# Organic Electrochemical Neurons: Nonlinear Tools for Complex Dynamics

**Gonzalo Rivera-Sierra[1], Roberto Fenollosa,[1] Juan Bisquert[1]\***

[1]Instituto de Tecnología Química (Universitat Politècnica de València-Consejo Superior de Investigaciones Científicas), Camino de Vera s/n, 46022, València, Spain.

*Email:* jbisquer@itq.upv.es

01 08 2025

**Abstract**

Hybrid oscillator architectures that combine feedback oscillators with self-sustained negative resistance oscillators have emerged as a promising platform for artificial neuron design. In this work, we introduce a modeling and analysis framework for amplifier-assisted organic electrochemical neurons, leveraging nonlinear dynamical systems theory. By formulating the system as coupled differential equations describing membrane voltage and internal state variables, we identify the conditions for self-sustained oscillations and characterize the resulting dynamics through nullclines, phase-space analysis, and bifurcation behavior, providing complementary insight to standard circuit-theoretic arguments of the operation of oscillators. Our simplified yet rigorous model enables tractable analysis of circuits integrating classical feedback components (e.g., operational amplifiers) with novel devices exhibiting negative differential resistance, such as organic electrochemical transistors (OECT). This approach reveals the core mechanisms behind oscillation generation, demonstrating the utility of dynamic systems theory in understanding and designing complex hybrid circuits. Beyond neuromorphic and bioelectronic applications, the proposed framework offers a generalizable foundation for developing tunable, biologically inspired oscillatory systems in sensing, signal processing, and adaptive control.

## Introduction

Emulating the spiking behavior of biological neurons using organic electronics has emerged as a promising avenue for neuromorphic computing and bio-interfacing applications. Among the various technologies explored, organic electrochemical transistors (OECT) based on organic mixed ionic–electronic conductors (OMIECs)[1-3] have attracted significant interest due to their intrinsic properties that enable complex signal processing and bio-compatibility. The integration of OECTs as the foundational element in biologically inspired spiking oscillators represents a significant and promising strategy for emulating the functional behavior of neurons [4-12] These devices play a central role in enabling perception-responsive systems that operate at the interface of ionic and electronic signaling, closely mimicking the way biological neurons process and respond to stimuli, and using similar physical principles to generate bio-compatible oscillations.[11,13]

A critical feature for achieving spiking dynamics in OECTs is a regime of negative transconductance, either through antiambipolar behavior resulting from partially stacked p–n heterojunctions in the transistor channel,[6] or through saturation effects in the density of states.[14] In previously reported architectures, OECTs have been configured into networks of multiple transistors, coupled with feedback-driven amplification stages, to replicate essential features of neural excitability and spiking.[6] However, the complexity of such multi-component systems poses challenges in integration, stability, and modeling.

In parallel, traditional oscillator theory has long provided analytical tools for understanding self-sustained oscillations in classical feedback-based circuits.[15-17] These include criteria such as those of Barkhausen, Nyquist, and Routh-Hurwitz, which analyze the loop gain, phase conditions, or characteristic equations of linearized systems to determine the onset of oscillations. Complementary time-domain analyses focus on the behavior of damping terms and the system's response to perturbations. These classical approaches are robust but rely heavily on the formulation of transfer functions and small-signal approximations.[18-21]

More recently, attention has shifted toward oscillators built from devices with emergent nonlinear physical properties, such as those exhibiting negative differential resistance.[12,22-25] In these systems, the interplay between the device's intrinsic relaxation dynamics and circuit-level capacitive elements forms an effective resonant system. Self-sustained oscillations[26] are maintained not by traditional feedback loops, but rather by the device's own nonlinear response—representing a fundamentally different mechanism for achieving oscillatory behavior.[27,28] In this context, the analysis of artificial neurons can be undertaken from widely used computational neuroscience models, that stem from the Hodgkin-Huxley model for the squid axon[29,30] and provide major tools to understand action potential dynamics and spiking.[31-33] These models rely on the framework of nonlinear dynamical systems, which are governed by differential equations involving time-dependent variables and fixed parameters.[27,34,35] Variables represent quantities such as voltage $u$ or internal currents that evolve over time, while parameters like capacitance or input current $I_0$ remain constant during operation.

A basic dynamical model is formed by coupled differential nonlinear equations,

$$\frac{du}{dt} = F(u, x, I_0) \tag{1}$$

$$\frac{dx}{dt} = G(u, x) \tag{2}$$

Here $x$ is an internal state variable and $F$ and $G$ are multivalued nonlinear functions. The nullclines represent the set of points in the system where a particular variable does not change over time, meaning its rate of change is zero. They are defined by the conditions

$$F(u, x, I_0) = 0 \tag{3}$$

$$G(u, x) = 0 \tag{4}$$

Dynamical systems theory[36,37] provides essential tools for understanding the long-term behavior of nonlinear systems such as (1, 2) by analyzing the asymptotic trajectories in phase space, which represent the evolution of system variables over time. These trajectories can converge to fixed points, diverge, or form closed loops—known as limit cycles—that correspond to sustained oscillations. By studying the stability of steady states, one can predict whether the system will remain at rest or transition to an active, oscillatory regime.

Crucially, even small changes in system parameters—such as ionic conductivity, gate voltage, or capacitance—can trigger qualitative changes in behavior. A central mechanism underlying the sensitivity to the occurrence of oscillations is the Hopf bifurcation, a phenomenon where a stable fixed point loses stability as a parameter crosses a critical threshold, giving rise to a limit cycle. This marks the onset of self-sustained oscillations from a previously quiescent state. In the context of artificial spiking neurons, including those based on organic electrochemical systems, the Hopf bifurcation provides a theoretical framework to explain how rhythmic spiking activity emerges from the interplay between system nonlinearity and feedback.[27]

Bridging these two worlds, feedback oscillators and self-sustained negative resistance oscillators, hybrid oscillator architectures have emerged as a compelling research direction. These systems combine feedback elements (e.g., operational amplifiers) with nonlinear devices like OMIEC-based OECTs, enabling the construction of compact, tunable, and biologically inspired spiking circuits.[6] Despite their potential, however, a comprehensive theoretical framework that captures the behavior of such hybrid systems—rooted in nonlinear dynamics rather than purely circuit-theoretic arguments—is still lacking.

In this work, we aim to fill that gap. We present a modeling and analysis framework for amplifier-assisted organic electrochemical neurons, using tools from nonlinear dynamical systems theory. By formulating the system as a set of coupled differential equations involving membrane voltage and internal state variables, we identify the conditions under which self-sustained oscillations emerge, and characterize them in terms of nullclines, phase-space trajectories, and bifurcation behavior.

In this way, it becomes possible to directly design and control the type of oscillations required for a specific application. While this work focuses on neuromorphic applications—particularly neuron-like behavior—the methodology can be readily extended to other domains, as it is grounded in general oscillator circuit principles. The universality of the underlying dynamics makes the proposed approach versatile and applicable across a broad range of technologies that rely on an oscillatory behavior.[38,39]

**Model**

A prominent example of a hybrid OECT oscillator has been reported recently by Harikesh et al.[6] The oscillator circuit of such hybrid oscillatory system is shown in Fig. 1a, where it is connected to a biological mouse model. As indicated in Fig. 1a, the neuron model is inspired by the combination of channels in the Hodgkin-Huxley model,[29,30] including both a potassium and a sodium channel, where the latter contains a negative transconductance feature that enables the sustained oscillations. In this setup, the system demonstrates a reproducible physiological response through the generation of autonomous spiking activity within a bio-relevant frequency range, all achieved with low operating voltages on the order of millivolts, Fig. 1b. This experimental result confirms the broader principle discussed above: the oscillator not only emulates key neuronal functions, but also interacts directly with living organisms, effectively integrating neuromorphic electronics with biological systems at the interface of bioelectronics and biophysics.

The implications for biophysical applications are extensive. However, in order to fully exploit the potential of such systems, both for biological integration and for high-performance and scalable technological platforms it is essential to understand the underlying circuit theory and the precise mechanism by which these oscillations are generated. Thus, in the following we develop a nonlinear dynamical approach that, to our knowledge, has not been shown previously.

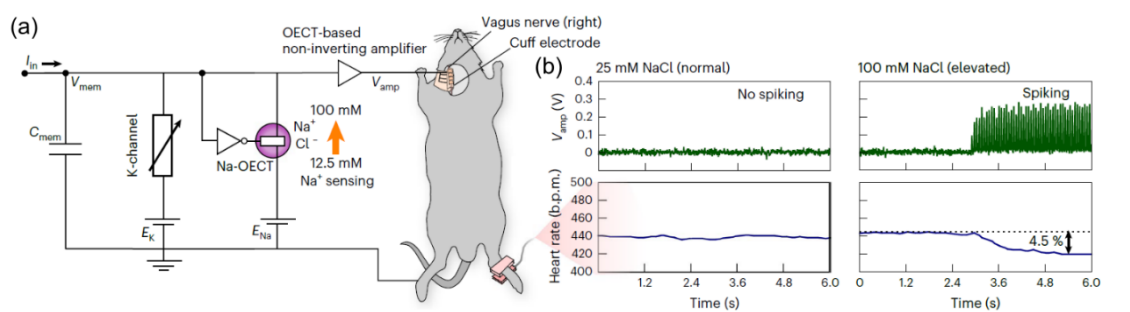

Fig. 1. (a) OECTs-based circuit showing sensing of Na+ ions at the Na-OECT and integration with the vagus nerve using an OECT-based amplifier and cuff electrodes. (b) Amplifier outputs at low (25 mM) and high (100 mM) concentrations of NaCl and the corresponding heart rate variation. The black horizontal dashed line represents the baseline heart rate. The purpose of this demonstration is to show the potential of OECTs-based oscillator to sense biochemical signals and interface with nerves and does not imply that new therapeutic means are developed. Reproduced from Harikesh, P. C.; Yang, C.-Y.; Wu, H.-Y.; Zhang, S.; Donahue, M. J.; Caravaca, A. S.; Huang, J.-D.; Olofsson, P. S.; Berggren, M.; Tu, D.; Fabiano, S., *Nat. Mater.* **2023**, *22*, 242-248, licensed under a Creative Commons Attribution 4.0 International License (CC BY 4.0).[6]

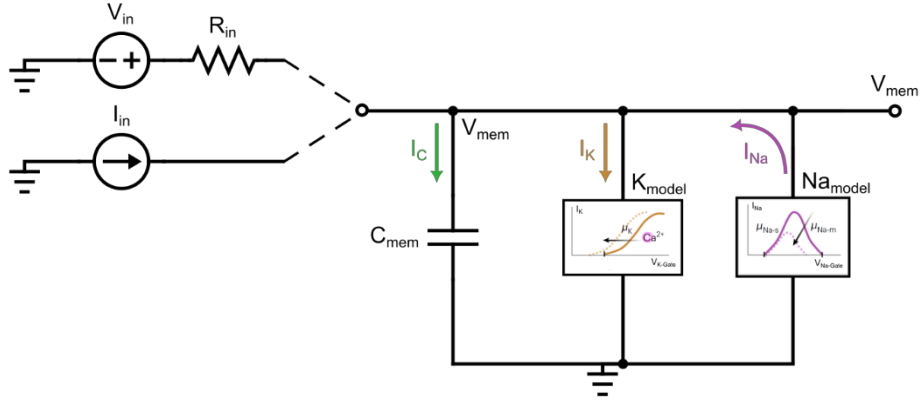

Fig. 2. Equivalent circuit schematic of the OECTs-based oscillator, inspired by the configuration shown in Fig. 1. The transistors characteristic I–V curves as a function of gate voltage $V_{Gate}$ are shown. It is assumed a quasi-static regime where the transfer function remains invariant with respect to the slow evolution of the membrane potential $V_{mem}$. The sodium ionic current channel contributes a feedback path to the rest of the circuit. The circuit can be operated either under a constant current input or using a constant voltage source with an input series resistor.

We analyze the differential equations governing the system of Fig. 1a by constructing an equivalent circuit (Fig. 2) and using models that accurately capture the behavior of the transistors involved. As a first simplification, we treat the potassium transistor branch by assuming a constant applied voltage $E_K$, and further neglect any significant modulation of the transistor current-voltage response due to variations in the drain-source voltage during oscillations. To account for such effects more rigorously, both the channel conductance $g_K$ and the activation variable $x$ would need to be expressed as functions of the potential difference $(V_{DS} = V_{mem} - E_K)$. This reduces its behavior to a typical resistive channel with a sigmoidal activation profile and an associated relaxation time:

$$I_K = g_K x V_{mem} \tag{5}$$

$$x_{eq}(V_{mem}) = \frac{1}{1+e^{-\frac{(V_{mem}-V_K)}{V_m}}} \tag{6}$$

$$\tau_K \frac{dx}{dt} = x_{eq} - x \tag{7}$$

where $x_{eq}$ represents the steady-state value of the variable $x$, $\tau_K$ denotes the relaxation time associated with the activation of the channel, and $V_K$ and $V_m$ correspond to the onset voltage and the steepness factor of the sigmoidal function, respectively.

Here, the parameters $g_K$, $V_m$, $V_K$ and $\tau_K$ vary depending on the experimental characteristics of the transistor and the applied activation voltage $E_K$. These must be properly adjusted to match the actual device response and reproduce the correct dynamic behavior in the model.

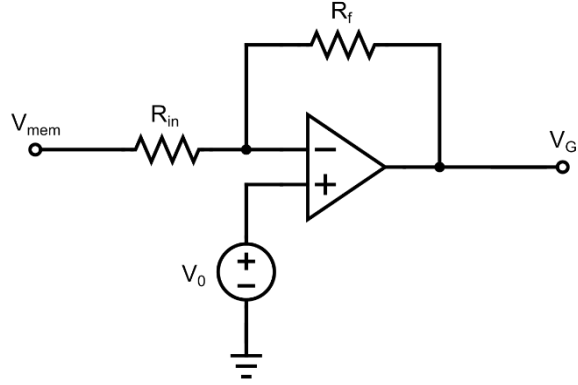

Fig. 3. Schematic of the inverting amplification stage implemented with an operational amplifier and a resistive feedback network. The membrane voltage $V_{mem}$ is applied at the input of the resistive divider, while the output of the operational amplifier provides an inverted and amplified signal that drives the gate terminal of the Na-OECT.

Now, turning to the analysis of the sodium branch, the first simplification arises from describing the voltage applied to the gate of the transistor as a function of the membrane voltage present in the circuit. To do this, we implement the simplest possible form of an inverting amplifier: a resistive network connected to the inverting input of an operational amplifier, while a baseline voltage is applied to the non-inverting input. By feeding the membrane voltage $V_{mem}$ into the resistive input network and connecting the amplifier output to the gate of the sodium transistor, we realize a complete inverting amplifier configuration, as shown in Fig. 3.

The relationship between the gate voltage and the membrane voltage is given by the amplifier gain $A$, which depends on the resistance values in the feedback network.

$$A = 1 + \frac{R_f}{R_{in}} \tag{8}$$

$$V_G = AV_0 - \left(\frac{R_f}{R_{in}}\right) V_{mem} \tag{9}$$

In this way, the inversion is essential to properly bias the sodium transistor in reverse, allowing its current to flow back into the rest of the circuit and provide the necessary feedback to sustain oscillations. Considering this configuration, the current flowing through the sodium branch, in the direction shown in Fig. 2, can be described as

$$I_{Na} = g_{Na} n(V_G) AV_0 - \frac{R_f}{R_{in}} g_{Na} n(V_G) V_{mem} \tag{10}$$

where $n(V_G)$ represents the activation and deactivation of the sodium channel. We describe this variable as a function of $V_G$ for simplicity and because of its direct relationship with the transistor's characteristic current–voltage curve. However, as previously discussed, this gate voltage directly depends on the membrane voltage. This $n(V_G)$ function could be modeled either as a Gaussian function with a maximum of 1, or as the difference of two sigmoid functions, similar to the one described in Equation (6), but with shifted threshold values. Accordingly, the internal activation variable of the sodium transistor was modeled as the difference between two shifted activation functions, as mentioned before: the positive branch corresponding to activation and the negative one to deactivation, following the expressions

$$n(V_G) = a(V_G) - d(V_G) \tag{11}$$

$$a(V_G) = \frac{1}{1+e^{-\frac{V_G - V_a}{V_{m_a}}}} \quad (activation) \tag{12}$$

$$d(V_G) = \frac{1}{1+e^{-\frac{V_G - V_d}{V_{m_d}}}} \quad (deactivation) \tag{13}$$

where $V_a$ and $V_d$ are the activation and deactivation, respectively, threshold voltages of the sodium channel, while $V_{m_a}$ and $V_{m_d}$ are the corresponding steepness factors leading the transition slopes. Depending on the specific experimental characteristics of the sodium transistor under study, both $n(V_G)$ and $g_{Na}$ can be tuned to accurately reproduce the transistor's response to gate voltage $V_G$.

In this analysis, we assume that the sodium channel activation voltage $E_{Na}$ is chosen such that the voltage difference between drain and source does not vary significantly within the membrane voltage range relevant to the oscillatory regime. If this condition is not met, then it becomes necessary to explicitly model the dependence of $g_{Na}$ and $n$ based on the drain-source voltage difference ($V_{DS} = E_{Na} - V_{mem}$).

Additionally, we consider the activation and deactivation of the sodium channel to be instantaneous, based on the assumption that its dynamics is significantly faster than that of the potassium channel. This allows us to simplify the model by avoiding an additional dynamic equation. However, for a more detailed analysis, one could include an additional differential equation analogous to Equation (7), but with a relaxation time for $n$ much shorter than that of the potassium activation variable $x$.

With these assumptions in place, the input current to the circuit is given by the following expression

$$I_{in} = C_{mem}\frac{dV_{mem}}{dt} + g_K x V_{mem} - g_{Na} n(V_G) V_0 + \frac{R_f}{R_{in}} g_{Na} n(V_G) V_{mem} \tag{14}$$

Combining this equation and Equation (7), we can now derive the two differential equations, $F(V_{mem}, x)$ and $G(V_{mem}, x)$, that govern the dynamics of the system in terms of the two state variables

$$F = \frac{dV_{mem}}{dt} = \frac{1}{C_{mem}}\left[I_{in} - g_K x V_{mem} + g_{Na} n(V_G) V_0 - \frac{R_f}{R_{in}} g_{Na} n(V_G) V_{mem}\right] \tag{15}$$

$$G = \frac{dx}{dt} = \frac{1}{\tau_K}\left[x_{eq}(V_{mem}) - x\right] \quad (16)$$

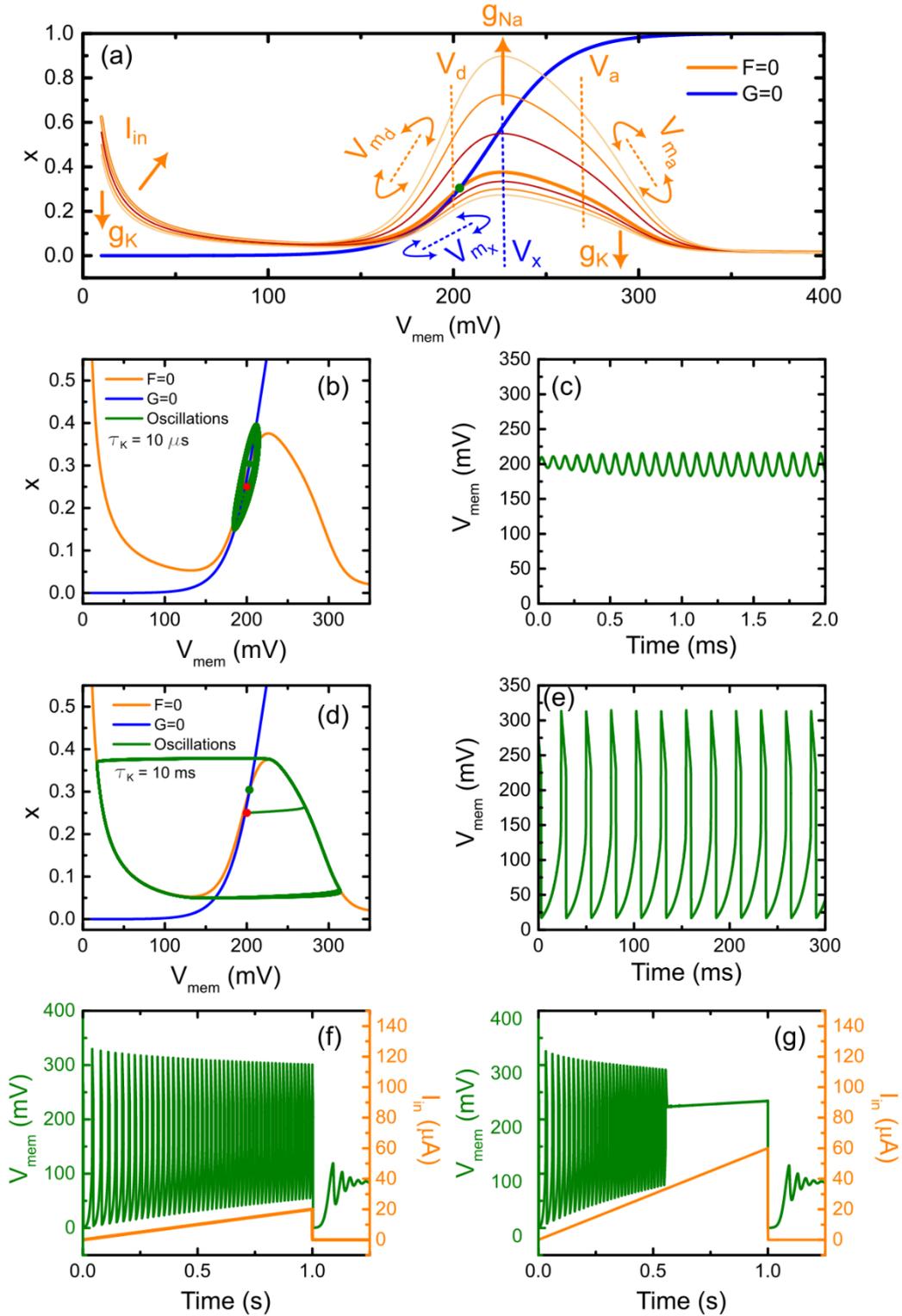

Fig. 4. (a) Graphical representation of the nullclines governing the system described in Fig. 1a and 2, plotted in the phase space ($x - V_{mem}$). The thick blue and orange lines

correspond to nullclines simulated with standard parameters: $C_m = 10\ nF$ ; $g_K = 0.80\ mS$ ; $V_x = 0.22\ V$ ; $V_{m_x} = 0.02\ V$ ; $R_f = 600\ k\Omega$ ; $R_{in} = 200\ k\Omega$ ; $V_0 = 0.35\ V$ ; $g_{Na} = 0.10\ mS$ ; $V_a = 0.50\ V$ ; $V_{m_a} = 0.04\ V$ ; $V_d = 0.80\ V$ ; $V_{m_d} = 0.04\ V$ ; $I_{in} = 5\ \mu A$. The upward-shifted nullclines result from an increase in sodium conductance: $g_{Na} = 0.15 - 0.25\ mS$, while the downward-shifted ones correspond to an increase in potassium conductance: $g_K = 0.90 - 1.10\ mS$. (b) Nullclines corresponding to the standard parameters and resulting system trajectory (green curve) for a potassium channel relaxation time of $\tau_K = 10\ \mu s$,. The red dot marks the initial condition, and the green dot marks the intersection point of the nullclines (c) Corresponding voltage time-domain oscillations. (d) and (e) Same as (b) and (c) but for a potassium relaxation time $\tau_K = 10\ ms$ . (f, g) Time-domain evolution of the system under increasing input current (right y-axis), illustrating the changes in the oscillatory behavior as this key parameter is varied.

**Discussion**

Once the two functions $F$ and $G$, which define the dynamic equations of the system's free variables, have been obtained, we can apply realistic parameter values for the transistors in order to analyze the equilibrium curves, the nullclines, the oscillatory region in phase space, and the emergence and disappearance of self-sustained oscillations accordingly. To illustrate the different physical behaviours attained by the system, we performed simulations based on the modeling assumptions discussed previously, and the results are shown in Fig. 4.

Using the nullclines derived from the equations $F = 0$ and $G = 0$, we can examine the oscillatory regime of the system. Oscillations occur when the intersection point of the nullclines is inherently unstable, specifically, when it lies within the negative resistance region of the sodium channel. In this configuration, that condition is met when the variable $x$ increases with increasing $V_{mem}$; that is, when the slope $\partial x/(\partial V_{mem}) > 0$ along the $F = 0$ nullcline. Therefore, if the $G = 0$ nullcline intersects the $F = 0$ nullcline in this region and only once, a limit cycle oscillation emerges. It should be clarified that although the mentioned slope is positive the differential resistance is in fact negative because the Na channel is considered to work in a reversed way.

To visualize this analysis, we refer to Fig. 4a. Here we show the shape of the system's nullclines and identify key parameters that influence their form. Among them, the input current stands out as the most accessible and impactful. The increase of this current modifies the shape of the $F = 0$ nullcline by increasing its decay at low voltages and narrowing the oscillatory region (i.e., decreasing the region of positive slope in $x$ at intermediate voltages). Other parameters—such as the activation steepness $V_{m_a}, V_{m_d}, V_{m_x}$, threshold voltages $V_a, V_d, V_x$, and intrinsic conductances $g_K, g_{Na}$—influence as well the nullclines depending on the experimental performance of the devices. Threshold voltages shift the oscillation range, steepness voltages affect the slope of activation/deactivation, and conductances shift the height of the potassium branch (lowering all the nullcline) or the sodium activation peak (raising the peak in the nullcline), respectively.

The externally tunable parameters in this system are limited to the input current and the activation/deactivation threshold voltages of the sodium and potassium channels, which can be adjusted through bias voltages as implemented in Fig. 1a. All other parameters are fixed by the experimental characteristics of the devices and the amplification conditions needed to accurately capture the sodium activation/deactivation window.

Figs. 4b and 4d display typical nullclines with a single intersection located within the oscillatory region, leading to the phase space trajectories shown in the same figures, and the corresponding temporal oscillations in Figs. 4c and 4e. The red and green dots indicate the initial and intersection points, respectively. In the first case (b and c), the potassium channel's relaxation time is short, $\tau_K = 10\ \mu s$, preventing the system from fully following the fast variable's nullcline ($F = 0$) and resulting in a compact limit cycle. In the second case (d and e), a longer relaxation time, $\tau_K = 10\ ms$, allows the system to jump along the fast variable's nullcline, producing larger amplitude oscillations (bigger limit cycle) driven by sharper transitions in the negative resistance region, close to relaxation oscillations. This difference in the dynamics produced by the decay time variation is, in fact, similar to what occurs in the saddle-node bifurcation, where the difference in the decay time yields either a typical saddle-node, characterized by more or less compact oscillations, or a saddle-node on invariant cycle, that features a large amplitude spike like oscillatory behavior.[40,41] This strengthens the point of view that apparently different neuromorphic systems are in a deeper way driven by similar differential equations.

Finally, Fig. 4f and 4g explore how varying the input current alters the oscillation regime. In the first case, the current is ramped from 0 to 20 $\mu A$ over one second and then abruptly dropped back to 0 $\mu A$. The system exhibits sudden emergence and smooth decay of oscillations. In the second case, the ramp reaches 60 $\mu A$. At this higher current, the $F = 0$ nullcline is sufficiently modified so that the intersection point exits the oscillatory region, suppressing oscillations. When the current decreases, oscillations reappear, again with decreasing amplitude. Notably, increasing the current leads to an increase in oscillation frequency and a reduction in amplitude—consistent with the shrinking of the oscillatory region in the nullcline space.

This oscillator circuit has also been recently applied in other contexts, such as in simple logic circuits that emulate neuronal operations through oscillatory behavior modulated by the mimic of voltage-gated dendritic calcium dynamics.[42] Furthermore, the negative transconductance exhibited by the OECT has been successfully exploited to develop single-transistor neuron architectures.[7,25] These advances highlight the critical importance of fully understanding, analyzing, and controlling the various types of responses that these emerging circuits can exhibit.

## Conclusion

In this work, we have developed a simplified yet rigorous dynamical model for analyzing hybrid oscillatory systems. By separating and describing the system in terms of its free variables, we enabled a tractable analysis of a complex circuit that combines classical feedback-based elements—such as operational amplifiers—with emerging devices exhibiting a negative differential resistance. Despite the apparent complexity of the circuit, this modeling approach allowed us to uncover the fundamental mechanisms responsible for the emergence of oscillations, highlighting the power of dynamic system analysis in capturing the essential behavior with minimal assumptions.

The theoretical framework presented here lays the groundwork for the design and control of a new class of hybrid oscillators. These compact architectures, which blend well-understood feedback mechanisms with novel device-level physics, are capable of producing self-sustained oscillations with high tunability and minimal component count. The use of phase space analysis and nullcline structure provides not only qualitative insight but also quantitative tools to determine how oscillations arise, how to control their amplitude and frequency, and which device or circuit parameters must be modified to meet specific functional goals. This level of control opens up new design possibilities for highly adaptable oscillatory systems.

Altogether, the approach outlined in this paper provides a robust and generalizable foundation for future work on oscillatory architectures—not only in the neuromorphic and bioelectronic domains, but also in broader fields where self-sustained oscillations play a central role, such as sensing, signal processing, and adaptive control. By grounding oscillator design in a dynamic systems perspective, we enable a new level of predictability and tunability, with potential to accelerate the development of next-generation hybrid electronic technologies.

**Acknowledgements**

This work was funded by the European Research Council (ERC) via Horizon Europe Advanced Grant, grant agreement nº 101097688 ("PeroSpiker").

**Declaration of Interest**

The authors declare no competing interests.

**Associated content**

Data Availability Statement

The code presented here can be accessed at https://doi.org/10.5281/zenodo.16677421 under the license CC BY 4.0 (Creative Commons Attribution 4.0 International).